# Subcritical insability of viscoelastic flow over a circular cylinder: A numerical study

Sai Peng[1,2], Jia-yu Li[1,2], Xin-hui Si[3], Xiao-yang Xu[4], Peng Yu[1,2,5*]

[1]Shenzhen Key Laboratory of Complex Aerospace Flows, Department of Mechanics and Aerospace Engineering, Southern University of Science and Technology, Shenzhen, 518055, China

[2]Guangdong Provincial Key Laboratory of Turbulence Research and Applications, Southern University of Science and Technology, Shenzhen, 518055, China

[3]Department of Applied Mathematics, University of Science and Technology Beijing,100083, China

[4]School of Computer Science and Technology, Xi'an University of Science and Technology, Xi'an 710054, China

[5]Center for Complex Flows and Soft Matter Research, Southern University of Science and Technology, Shenzhen, 518055, China

## Abstract

In this paper, we discuss whether the instability of viscoelastic flow around a circular cylinder is subcritical or supercritical by numerical simulation. The Oldroyd-B model is selected to describe the viscoelastic constitutive relationship. The Log-conformation reformulation is employed to stabilize numerical simulation. The parameter ranges investigated are the Reynolds numbers ($Re$) spanning from 5 to 100 and the Weissenberg number ($Wi$) spanning from 0 to 10, with a fixed viscosity ratio of $\beta$ =0.9. Simulations are performed under two paths, i.e., 1) increasing $Re$ (or $Wi$) and 2) decreasing $Re$ (or $Wi$) slowly and gradually from one state to the next. The results show that the statistical solutions such as the time-averaged velocity obtained along the two paths are not identical over certain parameter range, which is around the transition point from the steady to unsteady flow. This loading path dependence behaviour indicates that the flow instability is subcritical.

# 1. Introduction

Flow over a circular cylinder is a classic flow problem, which is often used as a benchmark case in numerical studies [1]. Over the range of the Reynolds number ($Re$) spanning from 5 to 100, the flow exhibits different different characteristics, depending on $Re$. The steady-state solution appears when $Re$ is less than 47 [2]. On the contrary, a linear instability grows and the flow develops into an unsteady state when $Re > 47$. As shown in Fig. 1(a), the transition from steady to periodic flow occurring at a ciritical point ($Re_c \approx 47$) is characterized by a supercritical Hopf bifurcation, which is independent of initial pertubation. Another classitical flow, Rayleigh-Benard convection [3], also exhibits supercritical instability. Supercritical instability is a local phenomenon, which could be alysized by linear stability analysis.

In contrast, there is also another kind of instability called subcritical instability [4], which is sensitive to initial disturbance. Fig. 1(b) shows a typical bufication diagram for subcritibcal instability. If an initial disturbance is below the dashed curve connecting the two ciritical points $A$ and $B$ ($A < B$), flow eventually evolves into a steady state. However, if the initial disturbance is strong (above the dashed cruve), flow eventually evolves into an unsteady state. Plane poiseuille flow [5] and vortex-induced vibration [6] exhibit subcritical instability. Subcritical instability need to be analyzed by nonlinear stability instead of linear stability. Flow with subcritical instability exihibits hysteresis, i.e., the lack of reversibility as cotrol parameter is varied. This indicates that the path dependence solution can be regared as a judgement cretrion for subcritical instability [7].

Viscoelastic fluid flow may exbibit more complex instability than that of Newtonian fluid. The governing equations for viscoelastic flow can be obtained by modifying the original N-S equations. Generally, a divergence term of elastic stress is linearly added to the right side of the momentum equations. The elastic stress needs to be modelled by a viscoelastic constitutive model related to a configuration tensor, such as the Oldroyd-B model [8], the FENE serial models [9], the Giesekus model [10], etc. Note that the FENE serial models, the Giesekus model and other complex nonlinear viscoelastic constitutive models are improved from the Oldroyd-B model. The configuration tensor transport equations can be described by an additional set of hyperbolic equations. Besides *Re*, another dimensionless parameter, i.e., the Weissenberg number (*Wi*), is often introduced to consider elastic nonlinearity in the dimensionless viscoelastic flow equations.

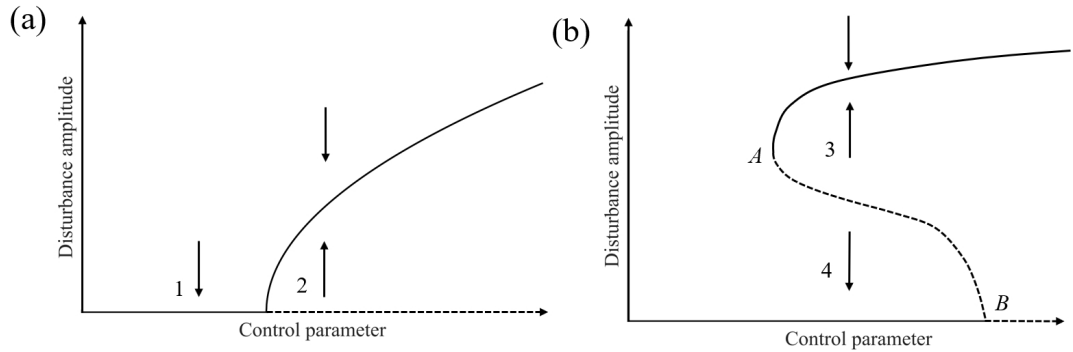

**Figure 1.** Buffication diagrams of (a) supercritical and (b) subcritical instabilities. Solid and dashed curves represens stable and unstable solutions, respectively. Routes 1 to 4 represens linear stability, supercritical instability, subcritical instability (finite-amplitude disturbance), linear stability (infinitesimal disturbance), respectively. Adapted from Guchenheimer and Holmes [4].

For viscoelastic flow at a low *Re* ($Re \ll 1$), convection terms in momentum equations can almost be ignored. However, at high *Wi*, due to nonlinear behaviour of elastic stress, viscoelastic flow may exhibit non-symmetric feature, unsteady behaviour, or even elastic turbulence with highly separated scales [11-13]. At moderate *Re* ($5 \leqslant Re \leqslant 100$), the influence of viscoelasticity on flow stability becomes more complex. For viscoelastic flow over a circular cylinder at low *Wi*, the viscoelastic effect stabilizes the flow. Correspondingly, the critical *Re* of the unsteady transition for viscoelastic flow at a low *Wi* may be higher than that for Newtonian flow [14]. However, at high *Wi* (e.g., *Wi* = 10), Peng *et al.* [15] showed that the viscoelastic flow at *Re* = 100 demonstrates much more complex instantanoues flow structures. This indicates that the complicated mechanisms including elasticity and shear-thinning may trigger other instabilities and the corresponding flow may become unstable.

In a low and moderate range of *Re*, Newtonian fluid flow past a single cylinder exhibits a supercritical transition. This is also true for the corresponding flow of viscoelastic fluid at low range of *Wi*. However, the rich flow phenomena revealed by Peng et al.'s numerical simulation [15] implied that instability type may be different at high *Wi*. At extraordinarily low *Re* but high *Wi*, the experimental studies on channel flow [7] and flow past side-by-side microcylinders in a confined channel [16] showed that the statistical behaviour induced by elastic instability exhibits initial value dependence. This indicated that the associated instabilility at high *Wi* is subcritical. Thus, it is reasonable to speculate that the instability in the high *Wi* viscoelastic flow past a cylinder is also subcritical. However, no theoretical, experimental, or numerical attempts have been made to verify this speculation. To shed some light on this matter, the present study conducts a series of symtematic simulations which consider the initial condition effect. The investigated parameter ranges are $10 \leq Re \leq 100$ and $0 \leq Wi \leq 10$. Note that in previous numerical studies, no initial value dependence behaviours were reported for viscoelastic flow past a cylinder. However, the initial conditions in these studies were often set to the corresponding Newtonian flow solution or an uniform flow field that meet the boundary conditions, while elastic stress was not introduced [15]. The initial disturbance introduced by these treatments may be too small to trigger the instability. Thus, the present study designs two paths to unveal the effect of the initial condition, i.e., 1) increasing *Re* (or *Wi*) and 2) decreasing *Re* (or *Wi*) slowly and gradually from one state to the next. The instability feature can then be identified by examining whether the corresponding paths for the statistical solutions such as the time-averaged velocity are exactly overlapped.

## 2. Formulation of the problem

The present study concerns a circular cylinder of diameter *D* immersed in a cross flow of a viscoelastic fluid with a uniform free stream velocity $u_{in}$. The governing equations of viscoelastic fluid flow consist of the incompressible continuity equation and the momentum equations [17-19]:

$$\nabla \cdot \mathbf{u} = 0, \tag{1}$$

$$\frac{\partial(\rho\mathbf{u})}{\partial t}+\nabla\cdot(\rho\mathbf{u}\mathbf{u})=-\nabla p+\nabla\cdot\boldsymbol{\tau},\tag{2}$$

where $\mathbf{u}$ is the velocity vector, $\rho$ is the density, $p$ is the pressure, $t$ is the time, and $\boldsymbol{\tau}$ is the total stress tensor, which can be decomposed into the solvent ($\boldsymbol{\tau}^s$) and polymer ($\boldsymbol{\tau}^p$) stresses, which reads,

$$\boldsymbol{\tau}=\boldsymbol{\tau}^s+\boldsymbol{\tau}^p.\tag{3}$$

In Eq. (3), $\boldsymbol{\tau}^s$ is determined by the Newtonian law $\boldsymbol{\tau}^s=\eta_s\left[\nabla\mathbf{u}+(\nabla\mathbf{u})^T\right]$, where $\eta_s$ is the solvent viscosity. Additional constitutive equations is needed to calculate the polymer stress tensor $\boldsymbol{\tau}^p$. In the present study, the Oldroyd-B model is selected as the viscoelastic constitutive relation, which can can be expressed as,

$$\frac{\partial\mathbf{c}}{\partial t}+\mathbf{u}\cdot\nabla\mathbf{c}=(\nabla\mathbf{u})^T\cdot\mathbf{c}+\mathbf{c}\cdot\nabla\mathbf{u}+\frac{\mathbf{c}-\mathbf{I}}{\lambda}.\tag{4}$$

The conformation tensor is defined as

$$\boldsymbol{\tau}^p=\frac{\eta_p}{\lambda}(\mathbf{c}-\mathbf{I}),\tag{5}$$

where $\eta_p$ is the polymer zero-shear rate viscosity, $\lambda$ is the relaxation time [20], and $\mathbf{I}$ is the second-order unit tensor.

Length, velocity, time, pressure, stresses, and vorticity are scaled by $D$, $u_{in}$, $D/u_{in}$, $\rho u_{in}^2$, $\rho u_{in}^2$, $u_{in}/D$, respectively. In order to characterize the present problem, a group of dimensionless parameters is used, including $Wi$, $Re$, and $\beta$. $Wi$ represents the ratio of the elastic to viscous forces [21]. $Re$ represents the ratio between the inertia and viscous forces. $\beta$ is the polymer viscosity ratio at vanishing shear rate, a measurement of polymer concentration and molecular characteristics. These dimensionless parameters are defined as follows,

$$Wi=\lambda\frac{u_{in}}{D},\quad Re=\frac{\rho u_{in}D}{\eta_0},\quad \beta=\frac{\eta_p}{\eta_0},\tag{6}$$

where $\eta_0=\eta_p+\eta_s$ is the summation of the polymer and solvent viscosities at the zero-shear rate.

## 3. Numerical method

The schematics of the computational domain and its dimensions are illustrated in Fig. 2. The origin of coordinates is placed at the center point of the cylinder. The distances between the origin and the inlet and outlet boundaries are $L_u$ = 17.5$D$ and $L_v$ = 50$D$, respectively. The vertical side length of the computational domain is $h$ = 25$D$. $X = x/D$ and $Y = y/D$ are the normalized $x$-coordinate and $y$-coordinate, respectively.

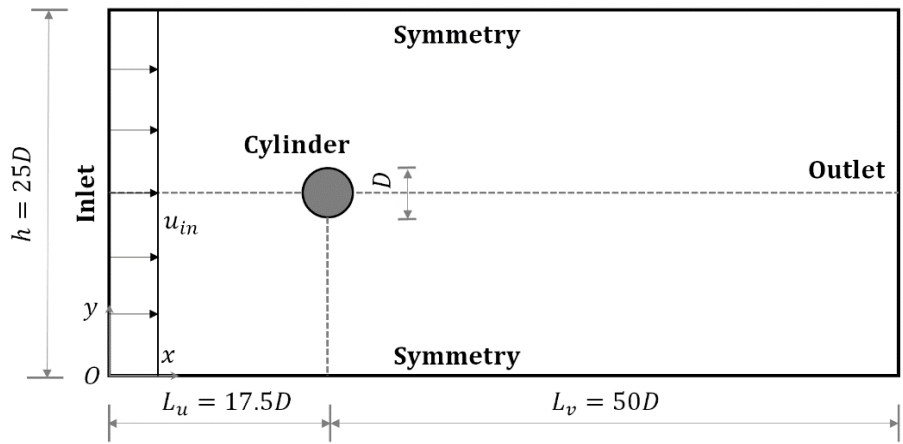

**Figure 2.** Schematics of computational domain.

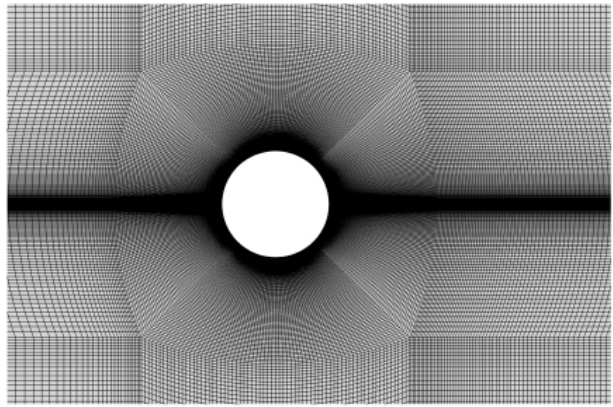

**Figure 3.** Details of the mesh near the cylinder.

The block-structured mesh is generated for this computational domain by using the commercial software ANSYS ICEM. As shown in Fig. 3, the surrounding region of the cylinder is discretized by the O-type mesh. The remainder of the computational domain is discretized using several blocks of rectangular meshes. The dense mesh is near the cylinder and the coarse mesh is near the domain boundaries. The domain and mesh sizes used in our previous work [15] are adopted in the present study.

In this work, the log-conformation tensor approach is adopted to stabilize the numerical simulation [22]. In the transformation, a new tensor $\Theta$ is defined as the natural logarithm of the conformation tensor. The governing equations are solved by the rheotool toolbox [24] intergrated in the open source CFD platform OpenFOAM [23]. The details related to the numerical method can be found in our previous work [15].

A uniform free stream velocity ($u_{in}$) is applied at the inlet boundary. A non-slip boundary condition is applied at the cylinder wall, whereas a symmetric boundary condition is imposed at the lateral boundaries. A zero-pressure condition is applied at the outflow boundary with a no-flux boundary condition for $\Theta$.

It is worth metioning that linearization technique is difficult to apply to the log-conformation tensor method, which poses great challenges on linear or nonlinear stability analysis of the present nonlinear viscoelastic flow stability problem. Alternatively, one may tackle this problem by introducing finite-amplitude perturbations in simulation and checking whether the flow can be driven to other stable solutions. These perturbations can be introduced when the simulation is initialized.

## 4. Results and Discussion

The control parameters considered herein are $Re$ and $Wi$, and two series of sumulations are performed correspodingly. When considering the effect of $Re$, $Wi$ is fixed at 10 while $Re$ is varied from 10 to 100 (Series I). When considering the effect of $Wi$, $Re$ is fixed at 100 while $Wi$ is varied from 0 to 10 (Series II). In all the simulations, $\beta$ is fixed at 0.9. Both series of simulations are symtematically performed under two paths, i.e., by gradually and slowly 1) increasing and 2) decreasing the control parameters (labelled by $IC$ and $DC$, repectively). The numerical solution of the current state is adopted to initialize the simulation for the next state. Near the transition points associated with large flow behaviour change, a smaller increasing (or decresing) step in the control parameter is adopted to ensure that any sudden jump can be carefully captured. In all the simulations, the velocity at the point (−0.6$D$, 0) in front of the cylinder is monitored and recorded.

First, the simulation results of Series I ($Wi$ = 10 and 10 ≤ $Re$ ≤ 100) are discussed. Fig. 4(a) shows the variation of the root mean square of the $y$-velocity ($v_{rms}$) at the monitoring point with $Re$ under two paths. $v_{rms}$ can partially reflect flow stability in front of the cylinder. For the paths $IC$ and $DC$, $v_{rms}$ is zero when $Re$ ≤ 22.5, which reflects the steady flow feature. For path $IC$, the steady state remains untill $Re$ up to ~39. Then an sudden jump in $v_{rms}$ occurs at $Re$ ≈ 39, which indicates that the flow become unsteady. $v_{rms}$ decreases with a further increase in $Re$. For the path $DC$, $v_{rms}$ increases with decreasing $Re$ over the range of 100 to ~40. There is an sudden drop in $v_{rms}$ at $Re$ ≈ 40, but $v_{rms}$ still maintains a relateviely high value over range of $Re$ from ~40 to 25. $v_{rms}$ then rapidly drops to zero with a further decrease in $Re$ to ~25 and remains zero if $Re$ ≤ 22.5. The results clear demonstrate that the two curves do not coincide between $Re$ = 22.5 and 40, which is around the intial unsteady transition point. The critical $Re$ for the unsteady transition occurs between $Re$ = 39 and 40 under the path $IC$, while this critical $Re$ is between 22.5 and 25 under the path $DC$. Both of these values are less than 47 of the Newtonian flow case. The sudden changes in both curves imply that the flow undergoes a nonlinear instability. The $v_{rms}$ – $Re$ curve is not reinstated on revseral of control parameter sweep, which indicates that the instability is subcritical.

The corresponding absolute $y$-velocity ($\overline{|v|}$) at the monitoring point for Series I is plotted in Fig. 4(c). As the monitoring point is located at the horizontal center line of the cylinder, a nonzero $\overline{|v|}$ indicates that the flow is asymmetric. For example, $\overline{|v|}$ is 0.049 at $Re$ = 30 for the path $IC$ and the corresponding time-averaged streamlines demonstrate obvious asymmetric flow structure in front of the cylinderas shown in the left bottom panel of Fig. 5(a). Also, a non-zero time-averaged $y$ velocity component $\overline{v}$ appears along the horizontal axis in front of the cylinder as shown in Fig. 6(a-i).

For the path *IC*, $\overline{|v|}$ changes slowly over the range of *Re* from 20 to 39, which means that the flow asymmetry holds. However, at $Re \approx 39$, $\overline{|v|}$ suddenly shoots up, which is consistent with the trend of the $v_{rms}$ – *Re* curve. Again, the sudden changes and the irreversible behaviour shown in Fig. 4(c) further confirm that the insability is nonlinear and subcritical. The flow exhibits multisability over the range of *Re* from 25 to 40. The time-averaged streamlines of *Re* = 40 are shown in the right bottom panel of Fig. 5(a). The flow asymmetry develops from upstream to downstream of the cylinder.

The simulation results for Series II ($0 \leq Wi \leq 10$, $Re = 100$) shows similar trends with those of Series I. Variations of $v_{rms}$ at the monitoring point with *Wi* is plotted in Fig. 4(b). For the path *IC*, $v_{rms}$ gradually decreases with increasing *Wi* for $Wi \leq 3.75$, e.g., $v_{rms} = 0.0127$ at $Wi = 0.5$ and $v_{rms} = 0.0017$ at $Wi = 2$, respectively. On the other hand, $v_{rms}$ rapidly increases with *Wi* over the range of *Wi* from 3.75 to 5.0. After that, $v_{rms}$ decreases with a further increase in *Wi*. For the path *DC*, the $v_{rms}$ – *Wi* curve does not coincide with that for the path *IC*.

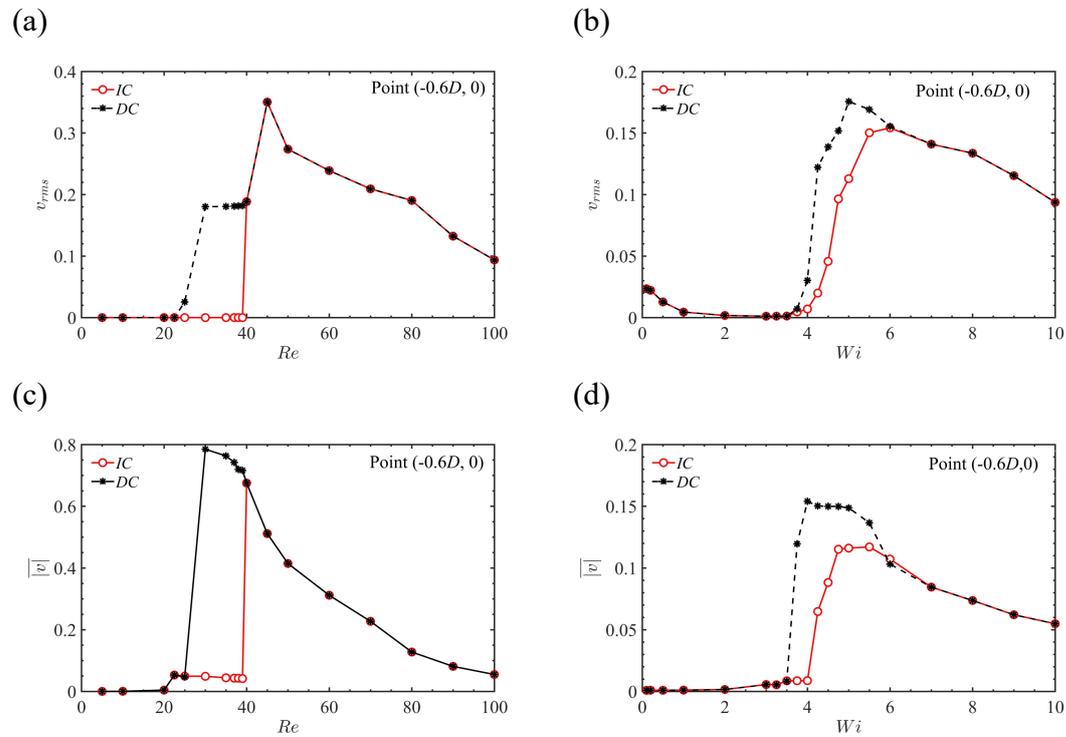

**Figure 4.** Variation of (a, b) $v_{lrms}$ and (c, d) $\overline{|v|}$ at the montoring point with (a, c) *Re* and (b, d) *Wi*.

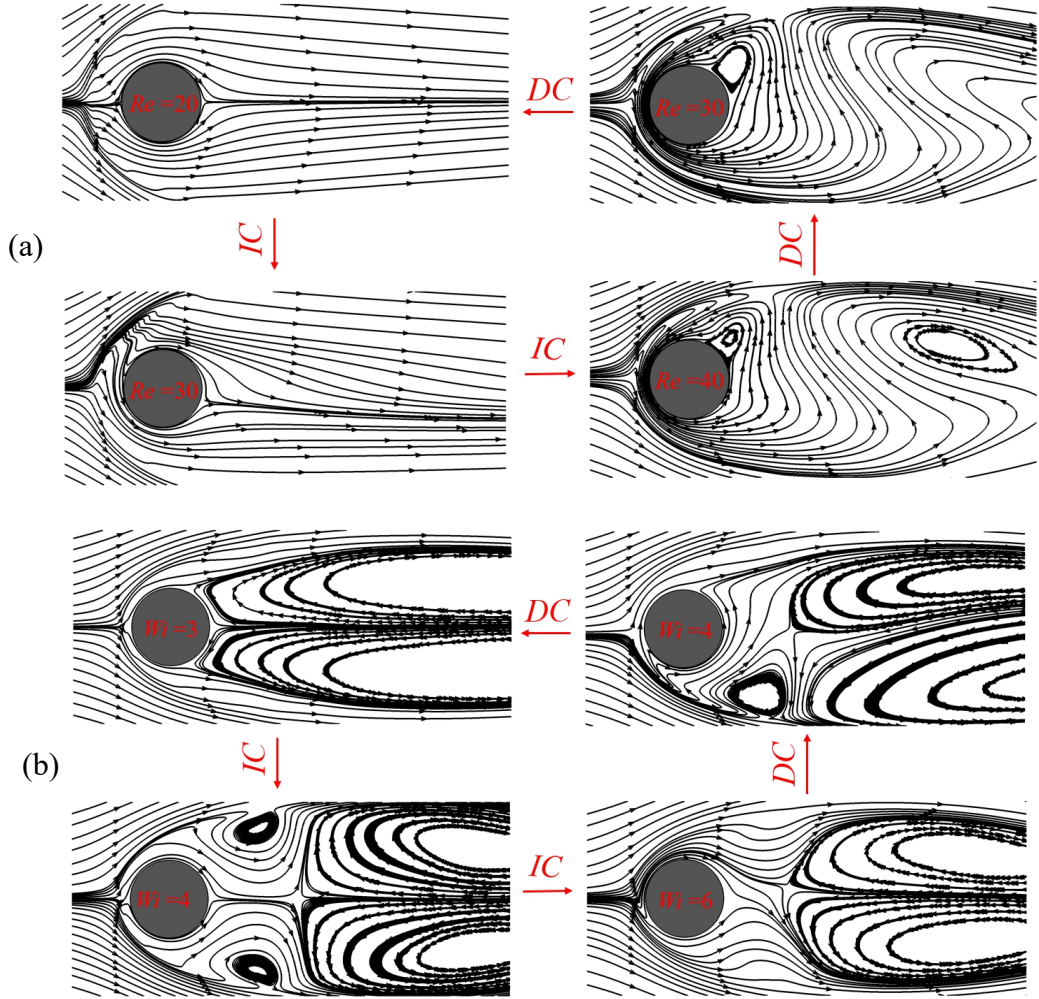

**Figure 5.** Time-averaged streamlines for (a) $Wi = 10$ and (b) $Re = 100$, and the other parameter is indicated in each panel. The arrows and symbols *IC* and *DC* denote the path for control parameter sweep.

Variations of $\overline{|v|}$ with $Wi$ is plotted in Fig. 4(d). For the path *IC*, $\overline{|v|}$ remains almost zero when $Wi \leq 2$ and slightly increases with $Wi$ over the range of $2 < Wi < 4$. $\overline{|v|}$ rapidly increases with $Wi$ over the range of $4 \leq Wi \leq 5$, which is constsient with that of $v_{rms}$. This means that with the intensification of flow instability, the asymmetry of flow field gradually appears. The asymmetry behaviour and the unsteady instability mutually promote each other. For the path *DC*, $\overline{|v|}$ always maintains at a high level over the range of $3.75 \leq Wi \leq 5$, while $v_{rms}$ is also very high. For example, $\overline{|v|}$ is equal to 0.154 at $Wi = 4$. At $Wi \approx 3.5$, a sudden change appears in $\overline{|v|}$, which is also accordance with that of $v_{rms}$.

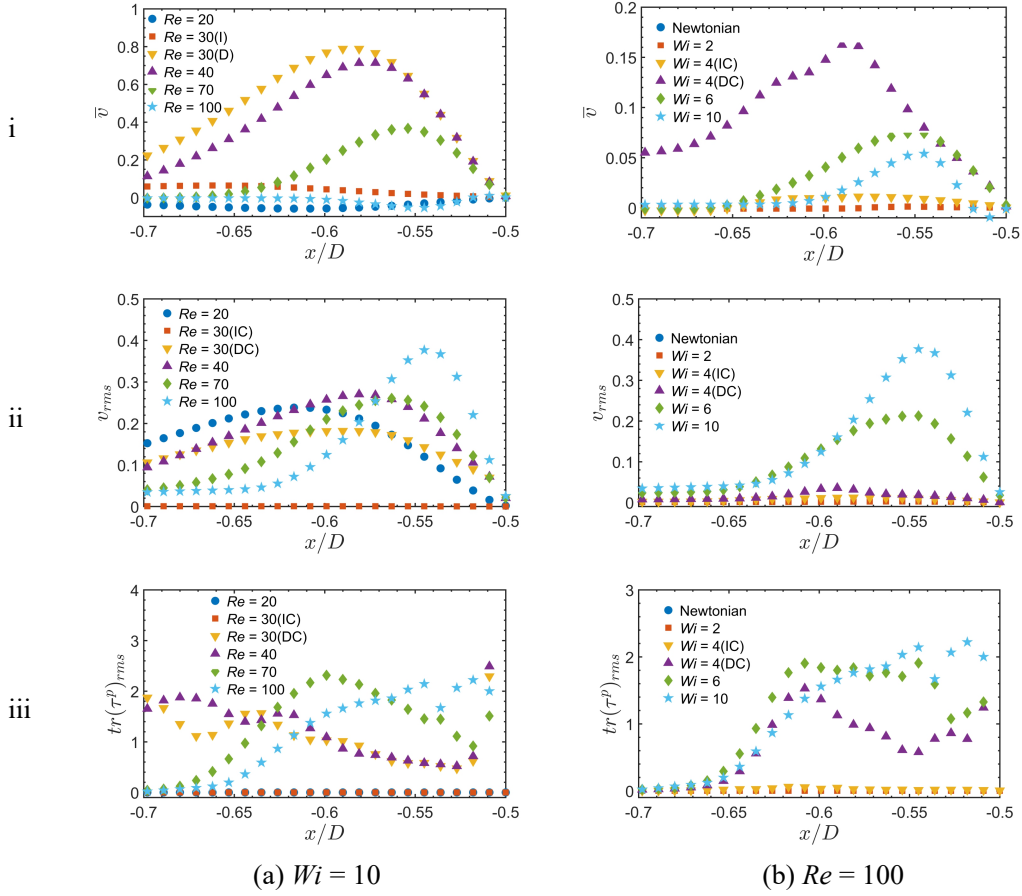

(a) $Wi = 10$            (b) $Re = 100$

**Figure 6.** Profiles of (i) the time-averaged $y$ velocity component ($\bar{v}$), (ii) the root mean square of the $y$ velocity component ($v_{rms}$), and (iii) the root mean square of the elastic stress tensor trace ($tr(\tau^p)_{rms}$) along the horizontal axis in front of the cylinder for (a) Series I and (b) Series II. The foremost point on the cylinder is located at $x/D = -0.5$.

Fig. 6(b-i) demonstrates that $\bar{v}$ along the horizontal axis in front of the cylinder is large at $Wi = 4$ (DC) and $Re = 100$. Correspondingly, the root mean square of elastic stress tensor trace ($tr(\tau^p)_{rms}$) is also large there as shown in Fig. 6(b-iii). However, the root mean square of vertical velocity ($v_{rms}$) at $Wi = 4$ (DC) and $Re = 100$ maintains a low level there as shown in Fig. 6 (b-ii). At high $Wi$ such as $Wi = 6$ and 10, both $v_{rms}$ and $tr(\tau^p)_{rms}$ are large near the leading edge of the cylinder when $Re = 100$, as shown in Fig. 6 (b-ii) and (b-iii). Because the Reynolds stress is proportional to $v_{rms}^2$, the magnitude of $tr(\tau^p)_{rms}$ in front of the cylinder is much higher than that of the Reynolds stress at $Re = 100$ when $Wi \geq 4$. This indicates that most of the flow energy

is in the form of the fluctuation of the elastic stress, which is called the elastic potential energy. The behaviours of $v_{rms}$ and $tr(\tau^p)_{rms}$ suggest that the flow fluctuation in front of the cylinder results from the fluctuation of elastic stress. The fluctuation of elastic stress is originated from the asymmetry of flow field. The flow asymmetry seems to accumulate the initial unstable energy induced by the viscoelastic flow instability. At the same time, the flow asymmetry is strengthened by flow instability. For the path *DC,* the flow asymmetry still appears both upstream and downstream at *Re* = 30, as shown in Fig. 5(a) and Fig. 6(a-i). The elastic potential energy stored in the flow asymmetry is slowly released with decreasing *Re* and the flow remains unsteady. When the flow asymmetry gradually fades to a certain degree and cannot hold sufficient amount of the elastic potential energy, the flow undergoes a rapid transition towards the symmetric steady state (although a very short period of the steady and asymmetry state exists).

For the viscoelastic flow at high *Wi*, the flow asymmetry and unsteady state characteristics exist simultaneously. Our numerical simulations indicate that the asymmetry and the unsteady behaviour show mutual reinforcing feature. For the path *IC* (increasing *Re* or *Wi*), the flow asymmetry gradually develops with the control parameter. When the elastic potential energy accumulates to certain amount, the unsteady tranistion is suddenly triggered, and strong asymmetric and unsteady behaviours appear. For the path *DC* (decreasing *Re* or *Wi*), due to the asymmetry of the flow, the accumulated elastic potential energy cannot be effectively released and the unsteady state persists. When the elastic potential energy is unloaded to a certain critical value, both the asymmetric and unsteady behaviours rapidly decay. In both paths, the elastic potential energy generated by the asymmetry of the flow has a certain hysteresis effect, which results in the irreversible behaviour on reversal of control parameter sweep.

## 5. Conclusion

In this paper, we examine the feature of flow instability for viscoelastic flow around a circular cylinder through numerical simulation. The Oldroyd-B model is selected to describe the viscoelastic constitutive relationship. Log-conformation reformulation is employed to stabilize numerical simulation. The control parameters considered herein include $Re$ ranged from 10 to 100 and $Wi$ from 0 to 10. The effect of the initial condition on the numerical simulation are investigated under two paths, i.e., 1) increasing $Re$ (or $Wi$) and 2) decreasing $Re$ (or $Wi$) slowly and gradually from one state to the next. The sudden changes of $v_{rms}$ and $\overline{|v|}$ indicates that the flow exhibits hysteresis and also multistability. The irreversible behaviour of $v_{rms}$ and $\overline{|v|}$ under the two paths indicates that the flow instability is subcrital. Both the sudden changes and the irreversible bahviour corresponds to the development of the nonlinear instability, which results from the accumulation of the elastic potential energy generated by the flow asymmetry.

## Acknowledgments

The author P Yu would like to thank the financial support from Shenzhen Science and Technology Innovation Commission (Grant No. JCYJ20180504165704491) , Guangdong Provincial Key Laboratory of Turbulence Research and Applications (Grant No. 2019B21203001), the National Natural Science Foundation of China (NSFC, Grant No. 12172163 and 11672124). The author XY Xu was funded by the National Natural Science Foundation of China (Grant No. 12071367), the Shaanxi Youth Top-notch Talent Program (Grant No. 289890259), and the Shaanxi Youth New Star Program of Science and Technology (Grant No. 2019KJXX-012). The author XH Si was funded by the National Natural Science Foundation of China (Grant No. 12072024). S Peng would also like to thank Dr. Zelong Yuan from Southern University of Science and Technology and Dr. Mengqi Zhang from National University of Singapore for useful discussion. This work is supported by Center for Computational Science and Engineering of Southern University of Science and Technology.